\input{aipcheck}

\documentclass[
    ,final            
  ]
  {aipproc}

\layoutstyle{8x11single}

\begin{document}

\title{An Overview of Meson-Nuclear Physics}


\author{Anthony W. Thomas}{
  address={CSSM, School of Chemistry and Physics, University of Adelaide \\
  Adelaide SA 5005, Australia}
}



\begin{abstract}
In this opening talk at MENU2010 we outline some of the key achievements in the field 
over the past few years as well as some of its major challenges and opportunities.
\end{abstract}

\maketitle


\section{Introduction}
This conference covers an extremely broad range of topics and in just 
a few pages it is impossible to even touch on all the areas which 
will be discussed. We have chosen to summarise just one area 
where there has recently been impressive progress, namely our 
quantitative understanding of strangeness in the nucleon. We also 
discuss a couple of examples where theoretical progress is of 
direct importance for future experiments as well as for the 
interpretation of hitherto anomalous results. Finally we make some 
remarks on meson and baryon spectroscopy and the exciting array of new 
facilities that are coming on-line in the near future.

\section{Strangeness}
The absence of strange valence quarks in the nucleon means that they can 
only contribute to hadron properties through the creation of virtual 
quark-anti-quark pairs and thus the accurate test of their role plays a 
role for QCD analogous to that of the Lamb shift in QED. Some 5 years 
ago the use of indirect techniques~\cite{Leinweber:1999nf}, involving lattice QCD, 
chiral extrapolation and measured baryon properties led to extremely 
accurate determinations of the strange magnetic moment and charge 
radius of the proton~\cite{Leinweber:2004tc,Leinweber:2006ug}. 
Just last year the magnetic moment calculation was 
beautifully confirmed by the first direct lattice 
calculation~\cite{Doi:2009vj}. Contrary 
to much speculation over the preceding decade, these contributions turn 
out to be quantitatively quite small - below 1\% for both the magnetic moment 
and the mean-square radius. The experimental results for the strange form factors 
of the proton~\cite{:2009zu,Acha:2006my,Aniol:2005zf,Maas:2004ta} are in 
remarkable agreement with the theoretical 
calculations~\cite{Young:2006jc}, 
although for now the errors are perhaps an order of magnitude larger.

Studies of hadron structure as a function of quark mass have provided a very 
natural explanation of these results~\cite{Thomas:2005qb}, 
namely that the finite size of the 
source of the meson clouds associated with chiral symmetry naturally suppress 
all Goldstone boson loops for meson masses above 
around 0.4 GeV~\cite{Young:2002ib}. The physical 
kaon mass falls above this limit and is naturally suppressed. The exploitation 
of this remarkable feature of nucleon structure through finite range regulated 
effective field theory has also led to a much more precise and again 
astonishingly small result for another property of strange quarks in the proton, 
namely the strange sigma commutator. Young and Thomas recently reported a value 
around 30MeV~\cite{Young:2009zb}, which is an order of magnitude 
smaller than commonly used and believed. 
This result is also in excellent agreement with direct calculations that appeared 
around the same time~\cite{Toussaint:2009pz,Ohki:2008ff}.

This new value for $\sigma_s$ not only has consequences for speculation about 
possible kaon condensation but surprisingly it is also extremely important for 
searches for neutralinos -- the favourite ``beyond-the-Standard-Model'' candidate 
for dark matter. Within the minimal supersymmetric extension it turns out that 
the biggest contribution to the neutralino-nucleon spin independent cross 
section came through $\sigma_s$, {\it until this new result}. As discussed by 
Giedt {\it et al.}~\cite{Giedt:2009mr}, the new value solves a long-standing problem of 
increasing the accuracy of predictions of these cross sections but it also 
leads to values at least an order of magnitude smaller than hitherto expected.
This has major implications for dark matter searches as well as for the interpetation 
of existing observations.
\begin{figure}
\includegraphics[width=3in,height=2.5in]{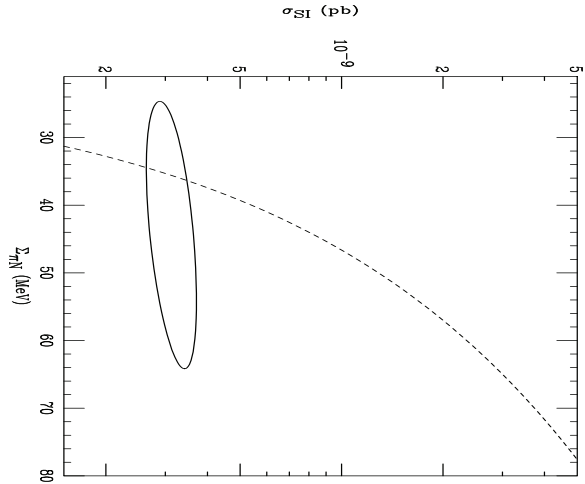}
\caption{Illustration of the improvement in the precision with which the 
neutralino-nucleon cross section can be calculated within a particular 
minimal supersymmetric extension of the Standard Model when the strange 
quark sigma commutator is taken from the recent analysis of lattice data 
by Giedt {\it et al.}~\protect\cite{Giedt:2009mr}, 
rather than through the $\pi N$ sigma commutator. 
Note that although the precision is much improved the absolute cross section 
is very much at the lower end of the range.}
\end{figure}

A final issue on which there has also been substantial recent progress concerns the 
parton distribution functions (pdfs) for strange quarks~\cite{Alekhin:2008mb}. 
This is extremely important in 
connection with the NuTeV anomaly~\cite{Zeller:2001hh},
where an asymmetry between $s$ and $\bar{s}$~\cite{Signal:1987gz} 
quarks can provide a potentially large correction. Global analyses of pdfs seem to indicate 
a positive value for $\langle x s^- \rangle$, but the level of precision needs 
substantial improvement. For this particular issue, lattice QCD seems unlikely to 
make a significant contribution in the near future. However, for the combination 
$\langle x (s \, + \, \bar{s}) \rangle$, the Kentucky group has recently been able  
to obtain quite an accurate value, namely $2.7 \pm 0.3$ \%~\cite{Deka:2008xr}. This may increase 
somewhat when the light quark masses take more physical values but the calculation 
nevertheless represents substantial progress.

\section{Spin dependendent phenomena}
There continues to be a great deal of effort devoted to clarifying the relative importance 
of quark and gluon spin and orbital angular momentum in defining the spin of the proton.
Data from RHIC has established that the total spin carried by gluons is  
less than about one half at a scale of a few GeV$^2$~\cite{:2008px,Abelev:2007vt}, 
and no doubt this will be pinned down even more 
precisely over the next few years. We already know that it is nowhere near large enough 
to play a significant role in resolving the famous spin crisis, where almost certainly 
it is the relativistic quark motion combined with gluon exchange and chiral symmetry 
that leads to the replacement of spin by quark orbital angular 
momentum~\cite{Myhrer:2007cf,Thomas:2008ga}. The debate 
over how best to define quark and gluon orbital angular momentum continues with 
vigour~\cite{Wakamatsu:2010qj,Chen:2009mr} and this will be a 
topic for discussion at this meeting.

Within the experimental community there is tremendous excitement over the host of new 
observables which promise to bring new light to the study of hadron 
structure~\cite{Belitsky:2005qn}. This 
includes a number of measurements involving semi-inclusive hadron production which 
will enable the determination of GPDs and TMDs~\cite{Afanasev:2007qh}. 
On the theoretical side, we note the 
tremendous progress made in lattice QCD to measure moments of the GPDs relevant 
to the proton spin problem~\cite{Bratt:2010jn}. 
Critical issues for the near future include the need to 
develop a deep understanding of the systematic errors associated with finite volume 
corrections and extrapolations in both pion mass and momentum transfer needed to 
accurately pin down the quark orbital angular momentum~\cite{Thomas:2010zz,Wang:2010hp}.

\subsection{Hadrons in-medium}
The nuclear community shows conflicting attitudes to the role of QCD in providing the 
fundamental explanation of the properties of atomic nuclei. For some QCD exists simply 
to give us chiral symmetry and nuclear structure can then be computed in terms 
of a number of low energy constants fixed by data. This is a more formal and systematic 
version of the older approach, still very much in favour, where one calculates nuclear 
properties using two-, three- and more-body forces, the former fitted to NN data while 
the latter are fit to chosen sets of nuclear data. The most modern expression of this
approach involves the use of the techniques of lattice QCD to compute the 
interaction of two and
three hadrons directly from QCD~\cite{Beane:2010em,Hatsuda:2009kq}. 
This is an ambitious long-term program, with current
calculations limited to rather small lattice volumes. Nevertheless, as we shall see 
at the conference some remarkable results have already been obtained.

An alternate approach begins with the realization that at some density 
(perhaps 3 to 10 times nuclear matter density) nuclear matter will
make a transition to quark matter -- a phase transition which may have dramatic 
effects on the observable properties of neutron stars. This suggests that one might 
begin to build a theory of the nuclear many-body system starting with a 
description of hadron structure at the quark level and then considering the 
self-consistent modification of that structure in a nuclear medium.
This is the approach taken within the QMC (quark-meson coupling) 
model~\cite{Guichon:1995ue,Saito:2005rv}. A remarkable
advantage of this approach is that no new parameters are needed to calculate the effective
density dependent forces~\cite{Guichon:2006er} between any hadrons 
whose quark structure is known. Indeed, it has
been possible to develop a remarkably successful derivation of realistic 
Skyrme forces~\cite{Guichon:2006er,Guichon:2004xg}
for comparison with low energy nuclear phenomenology -- while the fully relativistic
underlying theory successfully predicts key features of hypernuclear physics and allows the
study of the appearance of hyperons in dense matter~\cite{Guichon:2008zz}. 
Most recently, this model has also
been used to study the possibility of binding of charm-anti-charm mesons 
in matter~\cite{Krein:2010vp}, with
results very similar to those coming from direct 
lattice QCD simulations~\cite{Yokokawa:2006td}.
\begin{figure}
\includegraphics[width=3in,height=2.5in,angle=0]{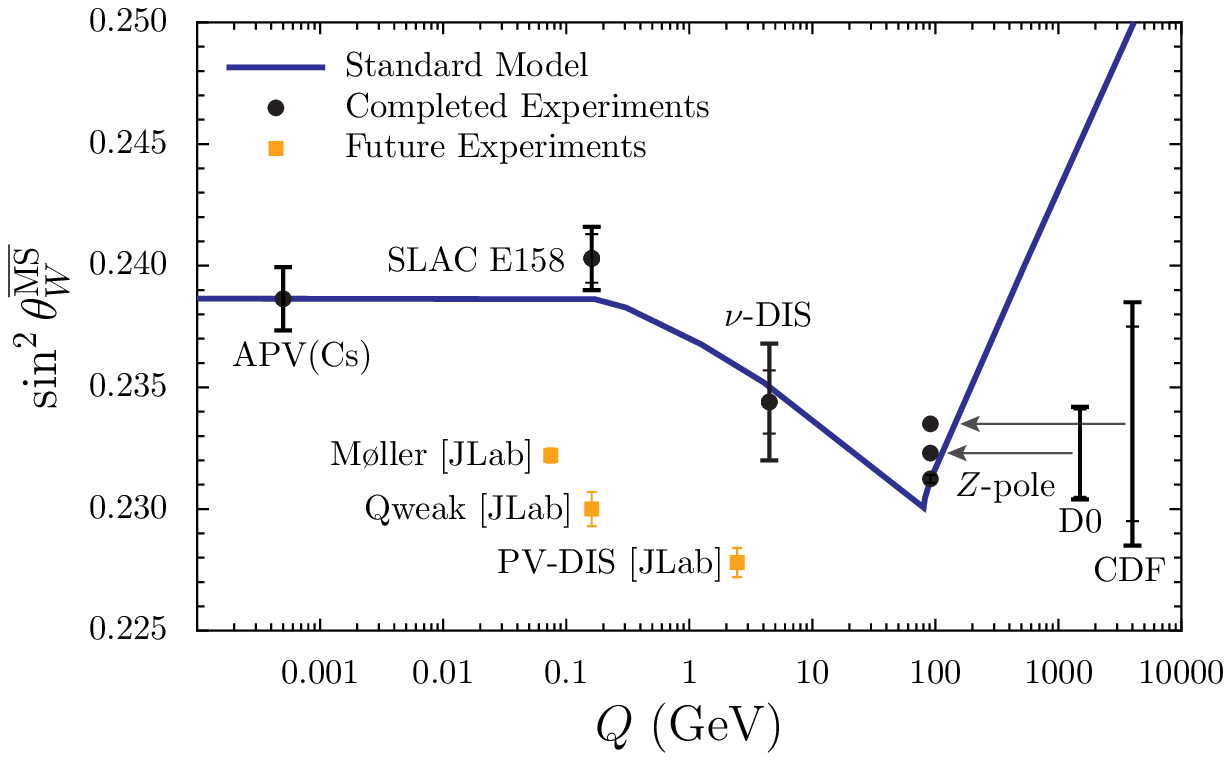}
\caption{
The curve represents the running of $\sin^2\theta_W$ in the $\bar{MS}$
renormalization scheme. 
The $Z$-pole point represents the combined results of six LEP and
SLC experiments, while the  
CDF and D0 collaboration results
(at the $Z$-pole) and the SLAC E158 result,
are labelled accordingly.
The atomic parity violating (APV) result
has been shifted from $Q^2 = 0$ for clarity.
The inner error bars represent the statistical uncertainty
and the outer error bars the total uncertainty
-- see Ref.~\cite{Bentz:2009yy} for details and associated references.
\label{fig:NuTeV}}
\end{figure}

The QMC model has the additional advantage that one can address not only those low energy
properties such as binding energies and charge densities but it can also be used to 
calculate the nuclear modification of the deep inelastic structure functions. 
Extensions of the QMC approach based upon a covariant, confining version of the NJL model  
have produced a satisfactory description of the EMC effect in finite 
nuclei~\cite{Cloet:2006bq}. It 
has also produced the remarkable prediction that should motivate quite a bit of 
experimental effort, namely that the EMC effect for the spin structure function 
should be roughly twice as large as the unpolarised 
EMC effect~\cite{Cloet:2006bq}. Another recent 
realisation of critical importance if one intends to use deep inelastic scattering 
data on nuclei in any way that assumes some sort of flavour symmetry, is that 
there will be a component of the EMC effect which is isovector in nature if one 
has a target with $N \neq Z$~\cite{Cloet:2009qs}. 
Most importantly, because the EMC effect involves a 
change in the structure of the bound nucleon, that isovector EMC correction will 
persist even if one derives data for an effectively isoscalar nucleus by subtracting 
the contribution of the excess neutrons. A very important example of this is the 
corresponding correction to the Paschos-Wolfenstein relation in the NuTeV determination 
of $\sin^2 \theta_W$. Figure~\ref{fig:NuTeV} shows the result of a recent reanalysis of 
the NuTeV anomaly~\cite{Bentz:2009yy} in which this correction 
was applied to the data along with a correction 
for genuine charge symmetry violation associated with the mass difference between 
$u$ and $d$ quarks.

\section{Hadron spectroscopy}
Many presentations at this meeting will address recent developments in 
hadron spectroscopy. The Excited Baryon Analysis Center at JLab is now 
producing detailed coupled channels calculations for two pion production 
as well as kaon production in pion and photon induced 
reactions~\cite{Sandorfi:2009xe,Lee:2009zzo}. In close 
collaboration with other groups at J\"ulich, Bonn and Mainz this has 
already clarified our understanding of some well established 
resonances~\cite{Nakamura:2010tc}. It 
has also established a sound basis for the extraction of information on any 
new states that may appear in the mass region above 1.8--2.0 GeV, where 
old quark model ideas suggest that there should be more states than have 
hitherto been seen. This coupled channel analysis is being challenged by 
a considerable quantity of new, high quality data that will help to answer 
these questions.

On the theoretical side, lattice QCD has again proven to be a critical player. 
In particular, the JLab group has remarkable new 
results for both charm quark~\cite{Dudek:2009kk} and light quark~\cite{Dudek:2009qf} 
systems, with a careful blend of group theory and lattice techniques now 
allowing the clear identification of a large number of excited states for each 
choice of spin and parity. As the masses of the light quarks in these 
simulations approach the physical values more closely it will be necessary to 
deal with effects of channel coupling here too.

\section{New facilities}
Across the world this field either has or soon will have the benefit 
of a large number of major new experimental facilities and there will 
be considerable discussion of their capabilities at this conference.
At this time the most noteworthy is, of course, JPARC in Japan, which 
only recently delivered its first beam. While the beam intensity is 
currently low and the energy currently 30 GeV rather than the 
ultimate 50 GeV the accelerator is already delivering a neutrino beam 
to the Kamiokande detector and the high momentum kaon line is ready for 
experiments. The first of those experiments, using pions rather than 
kaons, will take one more look at the infamous pentaquark. This 
object, which would add remarkably to our zoo of fundamental particles 
if confirmed, currently looks rather sick, with only one of the initial 
groups that reported a sighting still claiming to have evidence for 
its existence. Hopefully we will have news of that test at the next meeting.

Apart from its role in neutrino oscillations, the facilities 
at JPARC are ideally suited  
for extending our knowledge of hypernuclei dramatically. Of particular 
importance is the discovery of $\Xi$-hypernuclei, which are expected,  
within the QMC model discussed earlier, to be bound by bound by up to 
20 MeV~\cite{Tsushima:2009zh}. Establishing their existence and eventually the systematics of 
their binding and structure is not only important because it will establish 
a totally new area of study in nuclear structure but because of its 
importance for the properties of dense matter, as found for example in neutron 
stars. Certainly, if $\Xi$-hypernuclei are confirmed it will be difficult to 
avoid the conclusion that the $\Xi^-$ should appear at densities not far 
above $2 \rho_0$ in matter that is in $\beta$-equilibrium. Their appearance, 
in turn, has the effect of dramatically softening the equation of state and hence 
the observable properties of neutron stars.
Finally, we mention the promise of Drell-Yan for studies of the flavour 
structure of hadrons and nuclei as well as offering new types of spin-dependent 
measurements in which one can more easily detect processes which violate 
chiral symmetry.

Two future facilities, due to operate at a similar date, 
which will also undoubtedly be the source of intense 
interest are the 12 GeV upgrade at JLab (just a few kilometers away from 
this meeting) and the new 
facilities at GSI-FAIR. Hadron spectroscopy, especially the study of the 
existence and properties of the exotic mesons predicted within QCD, should 
be transformed by the Gluex experiment in Hall D. (Though we should also 
mention the important studies currently underway at COMPASS at CERN.) Critical 
to the success of this experiment, which should be able to access states up to 
as much as 2.6--2.8 GeV, is the dedicated 9 GeV photon beam with high linear  
polarisation for which Hall D was built, as well as the extremely high 
acceptance of the purpose built detector. GSI-FAIR  also aims to 
produce and study $\Xi$-hypernuclei and both new facilities, as well as JPARC, 
should be able to develop the spectroscopy of strange baryons to a new level. 

The development of a circulating anti-proton beam at GSI-FAIR will probably 
come a little after the opening of this new facility but the set of 
experiments already proposed for it, including precision charmonium spectroscopy, 
is very exciting.  
We leave the many other possibilities opened by these facilities for 
discussion at the meeting, except to emphasise the kinematic range opened for the study 
of the spin and flavour dependence of the parton distribution functions, as well as 
semi-inclusive deep inelastic processes by the 12 GeV upgrade.

We also expect that there will be lively discussion over the long term plans to build 
one or more electron-ion colliders in Europe and the United States. The very ambitious plans 
at BNL and JLab will come to a focus in the next long range plan for the United States, while 
the somewhat more modest plans at GSI-FAIR are still under development.
Certainly the plans for an electron-ion collider at BNL or JLab offer a tremendous 
opportunity to look for non-perturbative gluonic effects as well as allowing one 
to probe the spin and flavour structure of matter at a deeper level than possible 
before. One should also mention the possibility of installing an electron accelerator 
at CERN so that one could have electron-ion collisions at the LHC, the LHeC project.
This is extremely ambitious in terms of cost and energy reach (although it has 
no polarisation) but clearly would be motivated by discoveries at the LHC itself 
and therefore a decision must await the first discoveries there.

\section{Summary}
In this brief welcome I have been able to do little more than touch 
on a few of the exciting physics challenges that will be discussed in 
detail at this meeting, as well as very briefly highlighting the 
wonderful array of new facilities that will be crucial to the 
resolution of those challenges. I wish all participants the 
best for a productive meeting!


\begin{theacknowledgments}
This work was supported by the Australian Research Council through 
an Australian Laureate Fellowship and by the University of Adelaide.
\end{theacknowledgments}



\bibliographystyle{aipproc}   


%
\end{document}